# THE PERSPECTIVES OF DEMOCRATIC DECISION-MAKING IN THE INFORMATION SOCIETY

### Olaf Winkel

Professor of Public Management at the Berlin School of Economics and Law, Germany

## ABSTRACT

*In order to structure the debate on the democratic potentials of digital information technology Hubertus Buchstein in 1996 created three ideal types, net optimism, net pessimism and net neutrality. In this study the viability of these positions is put under scrutiny from a current viewpoint. As a result, it should be noted that all positions, however, contain both elements that are still viable today as well as components whose relevance now appears questionable. It should be emphasized that the digital information society, such as the net pessimistic side assumed, in fact has become in many ways a world of bad alternatives, in which it is increasingly difficult to bring legitimate yet conflicting interests to a common denominator. Particular clear this becomes in the increasing tension between the civic right to individual freedom and the state claim on interventions to safeguard social order. If one asks for how the today situation and the prospects for democratic decision making present themselves in the light of Buchstein's pure types, it will be recognizable that comprehensive and profound changes are in the making, but the contours of a future democracy are not yet discernible.*

## KEYWORDS

*Democratic Decision Making, Electronic Government, Electronic Democracy, Information Society, Information Technologies*

## 1. INTRODUCTION AND FUNDAMENTALS

In order to structure the discussion about the democratic potential of digital information technologies, Hubertus Buchstein introduced the ideal types of net-optimism, net-pessimism and net-neutralism in 1996, which reflect the core points of the views represented on this issue at the time [1]. Other authors took up his triad of terms and strove to develop them further [2].

Net-optimism stands for the assumption that the dissemination of digital information technologies involves improvements in the area of democratic decision-making already because of its specific nature. Net-neutralism is characterized by the belief that the effects of the informational technology change in this regard are ambivalent on the one hand, and on the other hand are malleable. Net-pessimism stands for a view according to which the democratic polity more likely accrues disadvantages instead of advantages in the transition to the digital information society due to specific socio-economic and immanent technical factors.

In the following chapters, the three ideal types and the related assumptions will be outlined in more detail and then put to the test from the current perspective. This is not only intended to evaluate the viability of the competing positions, but it should also provide the opportunity to gain new insights and suggestions for the solution of problems that a digital information society obligated to democratic values is confronted with today.

The considerations are indeed influenced by a European perspective, which for example finds expression in a high priority of data protection aspects, but the issues dealt with are essential and arise in all parts of the world, where basic social functions are shifted to electronic networks. Given the breadth and complexity of the research subject at hand, it is understood that such an investigation can only discover fundamental connections and trends. The basis of the considerations is a literature study enriched by Internet research, in which, in addition to specialist publications, non-scientific publications have been included for reasons of being up-to-date.

## 2. THE OPTIMISTIC POSITION

### 2.1 Core Statements

The net-optimistic position [3] assumes that the distribution of digital information technologies involves improvements in the area of democratic decision-making already because of its specific nature. It is insinuated here that network communication allows a universal access to politically-relevant information and processes, that it meets the one-way communication of mass media with two-way communication and that it is largely immune to authoritarian interventions.

With such a situational assessment, however, different versions of net-optimism are associated with different perspectives, whereby among others a market-oriented and civil society-oriented variety is distinguished between. In the market-oriented version, the optimistic situational assessment implies that the development of the Internet should largely be left to the market competition, which is attributed the capability of not only being able to best develop the economic, but also the political potentials of web communication. The civil society-oriented reading, like the market-oriented version, advises an avoidance of public authorities making direct interventions in the design of the web world, but does not assume a primarily market-shaped, but rather a primarily civic filling of the free spaces resulting there as the best way to develop the democratic potentials of digital information technologies.

### 2.2. Relevance from Today's Perspective

The following findings can be used as evidence of the viability of the optimistic position:
The number of Internet users has grown rapidly in the past two decades, as shown by the corresponding values for the Federal Republic of Germany [4]. An increasing number of women, less educated people, less affluent and older people have found their way onto the Internet. In underdeveloped countries in Africa, Asia and South America, which only have a patchy landline infrastructure and where poverty conflicts with the spread of personal computers, the proportion of the population still excluded from the Internet has been reduced on account of the potentials of the new mobile technologies[5].

The wealth of information with political relevance contained on the Internet has also grown tremendously.[6] In addition to state and municipal agencies, institutions from the civil society sector are among the most important information providers. Information brought about through whistleblowing that has revealed secret activities of security agencies or questionable schemes of business enterprises enjoys particular attention today.

The proliferation of the easy-to-handle instruments that are connected to the terms Web 2.0 and Social Media provide previously more passive users with the opportunity to participate actively in the network design [7]. "Enabling both identity expression and community building" [8], these innovations tend to blur the boundaries between private and public spheres and pander the communicational empowerment of population groups, which were previously limited to the receiver role and therefore doomed to be part of a silent majority, regarded as consenting with views of communicational elites [9].

With the advancement of the Internet, new forms of articulation of political opinions have also emerged, which are referred to as "low threshold participation" [10]. These include 'liking' political opinions and supporting online petitions as actions that only require very little effort and yet can still be effective if many join them. The related possibilities were recognized early on by non-governmental organizations [11], which use digital networks "to spread their interests to the public at the lowest possible cost and to mobilize supporters" [12].

The advantages that web 2.0 applications offer in the context of civic resistance have come into focus due to the so-called "Arab Spring" where these played an "important role as a multiplier, coordinator and amplifier of the protests" [13]. In Germany, the anti-nuclear protests that have recently flared up again and the persistent resistance to the new construction of the Stuttgart train station [14] have had a similar effect and thereby also uncovered the value of mobile devices, such as cell phones and smartphones, and services such as text messages and Twitter for the initiation and coordination of political actions. The app Berlin Against Nazis [15] is a progressive political project that is primarily based on mobile communication. It is supported by the Association for Democratic Culture in Berlin and is intended to inform its users about actions of the extreme right and to trigger and organize activities to counter them.

To the extent that civil society players have gained influence on the design of the political agenda, the possibilities of traditional mass media and the politicians associated with it to determine the topics of political discourse and to interpret them generally have diminished [16]. Not only has the communicative monopoly of broadcasting authorities loyal to the government been shaken in less established democracies such as Russia or Hungary, but in North America and Western Europe new constellations of power are also forming in this regard. This is not only due to the technically induced formation of alternative public spheres, but also due to interactions between the traditional mass media and Internet communication [17]. News broadcasts and television magazines can get engaged with the wishes of politicians less and less, as otherwise they run the risk of being overtaken by mailing lists, blogs or video posts and of losing credibility.

Where, as in Turkey, an authoritarian government attempted to deprive a civic opposition to the informational, communication and coordination opportunities of technical digital methods through the blocking of services, it can be seen that the technical features and the global structure of the electronic networks actually make such efforts largely ineffective [18].

Discursive procedures that are entirely or partially Internet-based are on the rise. The participatory budget [19], which was first tried out in Porto Alegre, Brazil [20], is among the widespread deliberative participation instruments. In the process, selected parts of the municipal budget planning are placed at the disposal of the citizenry. As the results achieved are non-binding, but rather have a recommendation character, this instrument does not affect the decision-making level.

The concept of Liquid Democracy, which provides fluid transitions between direct and indirect democracy and tends to favor grassroots decision-making against representative procedures, is applied in different political organizations [21]. Above all to be mentioned here are the Pirate Parties [22]. Their followers want Liquid Democracy not only to be realized internally, but also in society as a whole.

Insofar as the limits of economic rationality as a dominant logic of social organization have become evident and neoliberal policies have been criticized given the increasing social rejections, the doubt has grown about the positive political consequences of a development of the network world that is largely left to economic competition.

While the market-oriented version of net-optimism has lost credibility, the civil-society oriented reading was able to preserve its credibility. For advocates of such a position not only can refer to the growing importance of non-profit organizations and non-governmental organizations active on the Internet in the setting and interpretation of political topics and the implementation of campaigns in order to support their theses, but they can also refer to the role of civil society players in the maintenance of the organizational and technical infrastructure of the information society. To be mentioned here are the open source movement, the procedures for developing the digital encyclopedia Wikipedia or organizations, such as the Chaos Computer Club, which pursue exclusive network policy objectives. If Jeremy Rifkin should be right, the digital information society will in a later stage even develop into network-based production forms in which civil society and business cooperate as equal partners [23].

## 3. THE NEUTRAL POSITION

### 3.1 Core Statements

According to Buchstein, net-neutralism [24] stands for the conviction that information technology innovations are ambivalent in their effects on democratic decision-making, i.e. that they result equally in opportunities and risks. According to the ambivalence thesis, the opportunities lie where net-optimism emanates from primarily technology-induced changes. Unlike net-optimism, however, net-neutralism not only addresses the positive aspects of the information technology change, but also possible undesirable developments that show the flip side of the coin from its view.

It warns of new asymmetries in political participation as a result of a digital division of society, of a growing verbal aggression, of the spread of content that infringes upon basic democratic values, of a flood of information that can impair the political orientation of the citizenry and pave the way for populist and demagogic movements, of a loss of the political public, of the suffocation of participatory potential as a result of a comprehensive commercialization of the virtual world and of the identification and disciplining of communication participants with unwelcome political views.

From this perspective, the transition to the digital information society depends on working towards maximizing the opportunities and minimizing the risks of the information technology change through suitable decisions and measures. In addition to state and municipal authorities, stakeholders from business and civil society should also make contributions to this purpose.

### 3.2 Relevance from Today's Perspective

The fact that the authors whom Buchstein associates with the neutralist position also discussed the dangers that result for the democratic community from information technology innovations must today be regarded as a decisive contribution to the understanding of the relationship between democracy and digitization. Concerning the sustainability of the neutralist statements, the following should be noted:

The notion of a digital divide and resulting participation asymmetries is still plausible today as a counterpart to the vision of a society in which a universal access to knowledge bases, discourses and decision-making processes ensures for comparable participation opportunities. This is already true with regard to technically advanced countries. In Germany, for example, most, but still far from all citizens are integrated into Internet communication. And a lot of people who did not grow up with digital information technologies, but rather became familiar with them in a later stage of life, are quite perplexed by the immense dimensions of the virtual world and the possibilities of network communications, of which they only use a fraction.

With regard to underdeveloped countries, this problem is still far greater. Although the spread of mobile communications technologies in Africa and Latin America has definitely brought improvements, simply the illiteracy encountered there leads to a permanent exclusion of large portions of the population from the network world [25]. And where people have access via cell phones or smartphones, they are still far from holding the opportunities of a Central European household connected to the broadband network and outfitted with a local area network, that simultaneously not only has a PC and an Internet-capable TV, but also laptops and tablets at their disposal.

Indeed, new channels suitable for the broadening and deepening of the communication between political decision-makers and citizens have turned up, but intensive dialogues between representatives and represented are slow in coming [26].

Social segments, for instance educated people that already had more opportunities for participation than others early on, are over-represented on the Internet [27]. Apart from the fact that new information elites have emerged [28], well-known social asymmetries are therefore reflected in many areas of the virtual world.

Where the circle of contributors has expanded, the stigma of inferiority adheres to the low-threshold participation, which is insinuated by terms such as "slacktivism" or "clicktivism" [29]. Even the value of digital technically-based deliberations, which are mostly located on the local level, is controversial. In addition to the lack of a binding effect of results, the attack area is provided by the minor importance that such methods have compared to political processes, which take place at the national and international level without inclusion and partially even without the knowledge of the citizenry [30]. In this sense, with regard to the situation of European states, the question can be posed to what extent the statements made by citizens regarding minor parts of municipal budgeting can be considered significant as long as national governments and largely autonomous European institutions make fiscal decisions at the same time, which may impact the national budgets in the billions, and even in the event that this does not happen, still result in a serious redistribution of social wealth.

Warnings of a growing verbal aggression and the spreading of content that infringes upon basic democratic values have now proven to be more than justified. In addition to cyber-bullying and flaming to intimidate political opponents, phenomena of "cyber terror" have recently appeared [31]. So-called "cyber jihadists" [32] thus published the personal data of French soldiers and their families in social networks [33] after supporters of the terrorist militia Islamic State had presented videos of the murders of journalists and other hostages on YouTube [34].

It is clear that information technology innovations have not only led to an informational explosion, but also to an exponential expansion of communication options [35]. The question of whether this has caused orientation problems that favor populist and demagogic movements can only be speculated about. The influencing factors are too numerous here and the interdependencies are too complex. The assertion that such a relationship cannot simply be dismissed is supported by the "sheer volume of material, accounts and platforms" [36] of Islamic militants, right-wing extremists and other groups who propagate their ideologies on the Internet and by the interlocking of activities on the web with actions in the so-called real world, which range from flash mobs to mass rallies.

What has been available for the organization of democratic communities in modern mass societies has never been an all-encompassing public, but after all a network of overlapping partial publics [37], which has been dominated by television since the rise of this medium and that has been influenced by services such as websites and e-mail since the emergence of the Internet. It is obvious that this network has continuously expanded in the past and cracked [38]. The reason for this was above all the rapid differentiation processes in digital media, which are increasingly marginalizing print media. The most recent boost to the breakup of public space is resulting from the spread of the web 2.0. Insofar as former mostly passive users become active players in the virtual word, the "mixture of mail and forum services" is growing, which "combine public or public group communication with the possibility of exchanging personal messages" [39]. The problem is exacerbated by the fact that young population groups are increasingly turning their back to the offers of traditional mass media [40].

The scenario of a transparent society in which communication participants with unwelcome political attitudes can be identified, discriminated against and disciplined, which is described by Buchstein as "Panoptifizierung" [41], meanwhile has become concrete to a degree that many people even see the biggest danger therein for the democratic decision-making process. The fact that the loss of privacy has now become a feature of modern society is due to different causes. Firstly is the fact that more and more people have found access to an increasingly powerful Internet, which also means that more and more people can be intercepted and leave behind digital tracks, which can be evaluated and condensed into personality profiles[42]. Secondly, the rapid spread of mobile communication coupled with the landline communication now also allows for the creation of movement profiles [43]. And thirdly, the technical innovations are penetrating deeper and deeper into more areas of life [44].

Internet companies in particular have the ability to collect information about individual members of society and to connect to profiles. As long as these companies are limited to the pursuit of economic goals, then the effects of their position of power on the democratic decision-making process are still of an indirect nature. But the effects are direct where the companies work together with security agencies, such as intelligence services registered in the United States who themselves deal intensively with the acquisition and analysis of digital net communication [45].
If security agencies receive the ability to generate detailed personality profiles of more or less randomly selected citizens and implement wiretapping operations without serious cause and judicial approval, then the democratic decision-making process will not only be disturbed by this, but also fundamentally questioned. For privacy is a prerequisite of freedom and freedom is a prerequisite of participation. In established democracies, such as Germany, the identification of communication participants with unwelcome political attitudes may result in career disadvantages. In less developed democracies without a sound foundation that ensures the rule of law, far more severe sanctions threaten up to bodily harm and imprisonment.

The democratic decision-making process is not only at risk if intelligence services use their knowledge about individuals to manipulate their behavior or to pull them to account for their rebelliousness. It is already enough that political commitment in citizenry tends to be perceived as risky, because it can be brought to the surface at any time and may be punished [46].

The warning against a marginalization of the political potential of web communication as a result of the commercialization of the Internet has, on the other hand, proven to be unsubstantiated [47]. Although the economic importance that now belongs to the virtual world has likely exceeded the boldest expectations, the political utilization of the Internet has seen no end.

This, however, of course does not mean that commercialization would pass by the democratic decision-making process without a trace. In addition to the consequences that result from the mass collection of data for the purpose of profiling, other phenomena are to be considered here, which are more subtle and whose importance has only been rudimentarily researched. These include the fact that social networks and especially Facebook have a decisive say regarding which messages are disseminated on the Internet and therefore also regarding how the world is perceived by participants [48]. These include the fact that search engines and especially Google work as selecting filters and thus also contribute to the design of social reality without most of its users being aware of this [49]. And this also includes the fact that apps sourced by manufacturers such as Apple have a structured intervention into formerly open-design computer worlds and spaces of interaction [50].

As stated above, the assumptions of net-neutralism are not exhausted in the conviction that the opportunities offered by web communication for the development of the democratic community are faced with an abundance of dangers. It is also assumed that targeted interventions open up the opportunity to make opportunities productive and to manage risks. As regards this aspect, experiences collected in the past two decades have yielded an unclear picture. Some evidence suggests that such influences actually exist and it has already been possible to successfully utilize them:

The proposition that measures such as the development of broadband wiring, the support of uniform technical standards and data formats, the provision of basic components or the promotion of media literacy [51] have contributed to offsetting the proportions of off liners and on liners in modern countries within just a few years in favour of the latter category seems plausible. The fact that support programs have helped many people in underdeveloped countries, which were long excluded from web usage, to create moderate access via stationary and especially mobile devices [52] is likely also hard to deny. And the same goes for the fact that the rapid increase in the number of websites of governments, administrations, parliaments and parties and the willingness to make databases publicly available in the sense of open government [53] in principle counteract informational asymmetries.

The continued existence of a political sphere in an increasingly commercializing virtual world may well be related to state and municipal programs, which support web-based deliberation or to web policy successes, such as the defence of web neutrality [54]. The assumption that the decisive cause for the maintenance of political spaces on the Internet lies here will, however, not likely be supported by anyone.

In other areas, contrary to the net-neutral expectation, it has apparently not been possible to defuse the dangers to democracy, which emanate from network communication:

Verbal aggression and the dissemination of inhuman and anti-constitutional content up to the calling for genocide have become an integral part of the virtual world. Legal standards and state appeals, the willingness of commercial platform operators to cooperate with authorities to eliminate such content and also civil society projects, such as the development of "netiquette," an ethically sound code of conduct for the virtual world [55], could not prevent this. And so the resolution of the public, without which democratic decision-making in mass society is impossible, could not be avoided by a stronger commitment of broadcasting authorities and publishers on the Internet.

The inability to design network communication according to democratically acceptable rules is very clearly demonstrated also in the helplessness towards data-collecting companies and the intelligence services threatening privacy. Those who trace this helplessness back to the lack of adequate defence instruments are, however, embarking on thin ice. It is only clear that traditional, nationally-based control mechanisms of rulemaking, inspection and sanction of deviant behaviour are taking less and less effect in the virtual world [56].

Many people expect more success from so-called technical data protection, i.e. from efforts to integrate the safeguarding of privacy in technical solutions from the very start [57]. In principle, this is possible, because such systems are not determined by the hardware, but rather by the software, which does not follow a fixed technical, but rather a freely configurable social logic. As programmers can ensure that customers of an Internet company only receive access to their services if they give the company a detailed insight into their personal circumstances in return, they could also ensure that the collection of personal data is omitted and the saving of connection data is left out. Cross-border political activities and supranational regimes are considered as starting points to pave the way for multilateral security-providing software solutions [58].

In addition, it would be possible in principle to give Internet users the means with which they can protect themselves from the spying of sensitive information and communication relationships [59]. Electronic cryptography is the basic technology suitable for this purpose. However, this is mainly used in the version of the digital signature while the possibilities of encryption, i.e. confidentiality protecting encryption, remain in the background. Together with applications of anonymization and pseudonymisation, encryption today remains largely in the stage of procedures that are beyond the horizons of most users and are also partially difficult to handle. Demands that these technologies need to be "socially established and substantiated as legitimate concepts" [60] can rarely be heard. Elites in politics and economy, however, avail themselves of the benefits of such systems, for instance by switching to virtual private networks or commissioning tap-proof cell phones from special companies [61].

It should also finally be noted here that even in countries with a long democratic tradition, such as the United States of America or the Federal Republic of Germany, an effective control of intelligence services is missing, whereby the question remains as to whether this is not wanted or is simply not possible [62]. The momentum of intelligence services is not a new phenomenon, but it is gaining an entirely new dimension in a society in which more and more communication relationships are displaced onto electronic networks and therefore can be traced and intercepted. This finding seems to be particularly threatening given the recent developments in the field of artificial intelligence. While the National Security Agency, the largest intelligence service of the United States, according to media reports is already able today to record Internet traffic almost completely [63], they and other security agencies still lack suitable instruments to evaluate the large masses of data. The ability to do this could, however, soon be opened up by self-learning computer programs, which already play an important role in the economy today and, as a part of the discussion of big data, were unilaterally positively assessed not all that long ago [64].

## 4. THE PESSIMISTIC POSITION

### 4.1 Core Statements

The net-pessimistic position [65], to which Buchstein also assigns himself, is diametrically opposed to the optimistic and at a critical distance to the neutralist view. Regardless of whether corresponding effects are anticipated as more inherent to the system or as bound to the prerequisite of specific political interventions, from the standpoint of net-pessimism it falls short of assigning comprehensive democratic-conducive potential to information technology innovations. This position rather assumes that a broad transfer of political functions to electronic networks may even yield extremely negative effects. Negative developments that can be prevented through suitable measures according to the ambivalence theory, from the net-pessimistic view, become immediate and largely unavoidable consequences of the information technology change. In addition to asymmetrical socio-economic prerequisites, intrinsic technological factors are identified as the reason for this. The specific characteristics of digital information technologies on which the hopes of net-optimists rest make the world of networks into a world of poor alternatives in the eyes of net-pessimists, where an undesirable development in one area can often only be prevented by accepting an equally serious development in another. In such a situation, it only seems logical that net-pessimists desire a restrictive use of digital information technologies in political contexts. They want to see corresponding applications focused on the civil society sector and warn against any form of virtual democracy, which could be suitable for displacing conventional democratic decision-making procedures. They are therefore opposed to electronic elections and plebiscitary voting. Some authors who are mainly attributed to other ideal types share this view [66].

## 4.2 Relevance from Today's Perspective

As far as the retrospective assessment of the net-pessimistic position is concerned, the following findings and considerations have to be taken into account:

The rapid proliferation of net accesses in almost all regions of the world, the tremendous expansion of policy-related sources of information and communication options, the accompanying expansion of political and civil social fields of action, the relevatization of the dominant role of traditional mass media and the repeatedly proven resistance of web communication against authoritarian blocking attempts are among the phenomena, which are difficult to reconcile with a net-pessimistic view.

While the findings and interpretations suitable for substantiating the net-optimistic position tend to question the net-pessimistic view, the scratches and stains in the image of the virtual world that conveys net-neutralism also support the net-pessimistic position. This already applies to the fact that the scenario of a digital divide and an associated "democratic divide" [67] not only continues to have explosiveness with regard to underdeveloped societies, but also with regard to countries in North America and Europe when a distinction is not only made between access or non-access, but also by the type of access and by the intensity of use [68].

The flood of rudeness, obscenity and irrationality that has swept into the political discourse with the transition to the digital information society together with anti-constitutional and criminal content, and the inability to free the discourse of this, can be considered as further evidence of the validity of the net-pessimistic thinking. The same applies with regard to the orientation problems arising from the increasing complexity and their possible consequences, with regard to the gradual decay of a collective focus of attention, with regard to the questionable activities of companies such as Apple, Facebook and Google as well as with regard to intelligence services who accept privacy less and less and therefore undermine trust in the state and democracy.

Above all, it should be noted that the central net-pessimistic argument is still topical, according to which an increasing Internet-based society is transforming into a world of bad alternatives in many ways, where an undesirable development can often only be avoided by accepting another development that is just as undesirable. These already reveal the growing tensions between the requirements that arise from the increasing flood of information and the interests of maintaining a political public. While discourse platforms, electronic newspapers and group-specific used search engines counteract the danger that the increasing complexity of the virtual world is leading to a rise of populist movements, caused by a loss of orientation in citizenry, on the other hand they are also leading to a further disintegration of the public sphere, because they themselves belong to the driving forces of the medial differentiation. The question of how to have both at the same time – i.e. not only well-informed and politically capable of acting members of society, but also as a public sphere that can serve as the basis of democratic decision-making in mass society – thus continues to stand.

The fact that the progressive shift of social functions to electronic networks is allowing a society formerly determined by the both-as-well-as structural principle to become an either-or society is manifesting itself particularly clearly into an increasingly controversial relationship between the civil right to liberal freedom and the state claim to maintain order. While the investigative powers of the state's security agencies and therefore also the claim to power of the state was already limited due to technical restrictions in the past, a seamless monitoring of almost all areas of life and work became possible with the transition to the digital information society. Insofar as these possibilities are exhausted, an erosion of fundamental liberal rights threatens with fatal consequences for the democratic community.

At the same time, a technology exists in the confidentiality-protecting electronic cryptography that offers reliable protection against the spying of sensitive information and supports the assignment of pseudonyms and anonymization methods, through which communication records can be protected from the knowledge of third parties. The possibilities that exist here cannot only be seen in the current efforts of elites in politics and economy to protect against eavesdropping and surveillance with the help of specialized companies, but also in the experiences that were collected in recent years with the so-called "deep net" [69] or "dark net" [70]. These terms stand for a segment of the Internet that can be used anonymously with the help of a special browser, through which you can, according to reports by journalists, even procure "drugs, weapons, killer services and child pornography" [71].

So there is not only the world of conventional browsers, search engines and platforms on the Internet where transparent customers become the norm for corporations and transparent citizens become the norm for security agencies, but a "parallel network" offering an "anonymity of design" also exists, before which "even intelligence services capitulate" [72]. In addition to the scenario of an Orwellian surveillance society, a radically liberal anarchy is therefore also conceivable here, in which the only applicable law is that of the strongest. What has not been apparent so far is the road to a digital information society in which civic freedoms and state claims for order subsist next to each other as constituent elements of a democratic polity respectively exist in a balanced relationship to each other.

Demands for electronic elections, sometimes also for electronic plebiscitary polling, were already heard in the mid-nineties and still have not fallen silent. An argument frequently presented for the introduction of online elections is that these are suitable for counteracting the "voting abstention of young population groups" [73]. As represented above, however, net-pessimists assume that information technological innovations are only suitable for the lower levels of the democratic decision-making process. The following should be noted in this regard:

The demand or expectation that electronic elections and voting should not play a role was almost universally met by the developments of the last twenty years. Even the Federal Republic of Germany has not proceeded past the occasional experiment with electronic elections, whereby these were usually still outside of the governmental [74].

Deviations from this trend can be found in Estonia, where citizens have been able to vote in local elections, European elections and parliamentary elections on the Internet [75]. A "top spot" is occasionally conceded in Brazil [76]. However, it should also be noted that there the focus has not been placed on the online election, but on the use of electronic voting machines [77]. In Switzerland, even plebiscitary voting took place on the web [78].

It is worth noting here that the rejection of information technological innovations at the decision-making level is no longer legitimized, like it was twenty years ago, by warnings against a "push button democracy" [79] stripped of all power of symbolism and "push button voting" [80] in which faceless masses make questionable decisions, but by references to security problems [81]. The fact that such arguments are not easily refutable is already owed to the truth that absolute security is an illusion, even in the virtual world of networks.

# 5. CONCLUSIONS AND FINAL CONSIDERATIONS

Even twenty years after Hubertus Buchstein structured the discourse on the democratic potentials of digital information technologies by introducing the ideal types of net-optimism, net-neutralism and net- pessimism, no position can be universally confirmed or refuted. Rather, it is clear that all viewpoints have elements on the one hand that still appear substantial even from today's perspective, but on the other hand also contain components whose viability is questioned by developments in the last two decades. The fact that the strengths of one position point to the weaknesses of the other position and vice versa is the nature of the matter.

The fact that the balances of power in political communication have shifted in favour of the citizenry and the rule of the public by traditional elites has largely slipped from their hands can be cited as a weighty argument to confirm the optimistic view. Neutralist warnings of a brutalization and rationalization of political discourse, of a breakup of the political public and of the escalation of monitoring as well as network pessimistic indications that the intervention possibilities for curbing such phenomena are limited seem almost prophetic today. The latter can be seen in the context of the finding that a society, which is increasingly shifting social functions to electronic networks, tends to turn into a world of bad alternatives, in which it becomes more and more difficult for legitimate, yet conflicting interests to be brought to a common denominator. This becomes particularly evident in the increasingly conflictal relationship between the civic right to individual freedom and privacy on the one side and the state claim to order on the other. The maxim, by which the neutralist view deserves the advantage in cases of doubt since the belief in configurability that it conveys is an indispensable prerequisite for successful design [82] should still apply despite the design restrictions that have come to light in recent years.

If the perspective is changed and the viability of the different positions is not asked about, but rather how the current and future challenges are represented in light of Buchstein's triad of terms, this only results in few concrete findings. It is only clear that the democratic decision-making process of the future will look different in many respects from the way it used to be in the past. Overall, more questions arise here than those that are answered, which were posed by the authors whom Buchstein had in mind when developing his ideal types.

The new questions include the following:

What consequences arise for democratic decision-making from the fact that not only parts of the population remain excluded from the advantages of network communication, but another divide is emerging because the one division is communicating in an increasingly safe and confidential manner, while the other is exposed to the continuous access of third parties?

How should a democratic polity deal with phenomena, such as the spread of inhuman content, which is contrary to its fundamental values on the one hand, but on the other hand is beyond its reach? Is it better here to insist on restrictive laws to at least denounce deviant behavior, or should the state hold back in this regard in order to not let its own powerlessness be revealed all too clearly?

What effects does this have on democratic processes if Internet companies, such as Google and Facebook, have significant and partially also covert influence on how the world is perceived in large parts of the population?

How does the democratic decision-making change if, on the one hand, privacy can no longer be guaranteed, because intelligence agencies and Internet companies are always recording information and communication relationships and using or misusing them for their own purposes, but these incidents on the other hand are also very likely discovered through whistleblowers and platforms such as Wikileaks and become the object of political discourse?

Do innovations, such as digital technology-supported participatory budgets, pave the way for a democracy more determined by civil society, or are they to be interpreted less as an offer for power sharing and more as symbolic gestures that distract from the fact that citizens are not empowered, but rather are disempowered given the progressive political interweaving at the supranational level and powerful economic imperatives?

Are the security concerns about electronic voting actually owed to technical restrictions or are they merely advanced in order to get a political discourse about the enrichment of the representative democracy through direct democratic elements, which is demanded by different sides, out of the way?

Are there options in addition to the introduction of electronic elections and voting to make the information technology change productive for the development of democratic decision-making? Where are the opportunities and risks of instruments through which citizens are directly incorporated into the legislative process, and of innovations that rely on a combination of directly democratic and deliberative procedures? Is the concept of Liquid Democracy failed or is it yet able to open up new perspectives for the design of democratic processes not only within political organizations, but also to society as a whole?

Answers to such questions are still characterized as being in outlines at best. But even if they could already be given today, the perspectives of modern democracy would not yet be fully illuminated by that. For in addition to socio-technical factors, socio-economic and associated regulatory political factors are also crucial in this regard. So it makes a big difference for the continued development of modern democracy whether the "contingent hegemony of neoliberalism" will continue to intensify and define policy in the role of an auxiliary agent of economy [83], whether "democracy will be able to hedge in the economy" and thus can regain its ability to control [84], or whether, as predicted by Rifkin, a "decentralized and collaborative industrial revolution" [85] will lead to new forms of economic activity in which capitalistic and civil society-based methods of production will complement each other equally and social and political balances of power will shift accordingly.

In the area of political decision-making, comprehensive and profound changes are in the making, whereby the contours of a future democracy are not yet recognizable.

## AUTHOR

Dr. Olaf Winkel is Professor at the Berlin School of Economics and Law. He obtained a master's degree in Sociology and doctorate and post-doctoral lecturing qualifications in Politics. Previous stations in his professional life were among others the Universities of Münster and Bochum (Germany). As an expert for Electronic Government he also deals with the perspectives of democratic decision-making in the Information Society.

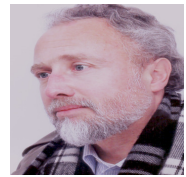